# *FERRON TYPE OF CODUCTIVITY IN METAL* $CuFeSe_2$

**A. M. Polubotko**

A.F. Ioffe Physico-Technical Institute Russian Academy of Sciences,

Politechnicheskaya 26 194021 Saint Petersburg RUSSIA

E-mail: alex.marina@mail.ioffe.ru   Tel: (812) 247-99-73,

Fax: (812) 247-10-17

**ABSTRACT**

It is pointed out, that in the compound $CuFeSe_2$ the charge transfer in the paramagnetic region is carried out by ferrons of a small radius predicted by N. Mott. For very low temperatures the charge transfer may be carried out by ferrons of a large radius. The results are well confirmed by the temperature dependence of resistivity and by metallic type of conductivity of $CuFeSe_2$ and are also well coordinated with the character of charge transfer in magnetic materials.

Investigation of $A_1B_8C_6^2$ compounds is of great interest. Two of them $CuFeS_2$ and $CuFeTe_2$ have unique properties, as it was found in [1-2]. They are zero-gap magnetic semiconductors with a ferron character of charge transfer*. One can expect, that their isoelectronic analogues have similar properties. Recently investigation of $CuFeSe_2$ was carried out [4]. In opposite to $CuFeS_2$ and $CuFeTe_2$, $CuFeSe_2$ has another crystal structure. It crystallizes in the eskebornite lattice. In [5] the magnetic structure of $CuFeSe_2$ was investigated. In accordance with results of these investigations $CuFeSe_2$ transfers to ferrimagnetic state at $T_c < 80$ K. Its crystal cell has parameters $(a,a,c)$, while its magnetic sell is doubled along one of axes and has the parameters $(2a,a,c)$. A mean magnetic moment of iron atoms is ~1.75 $\mu_B$ with the direction ~$15^0$ with respect to the y-z plane. The magnetic moments, which are situated in various crystal sites differ one from another on the value ~0.03 $\mu_B$, that causes weak ferrimagnetism. Temperature dependencies of resistivity, magnetic susceptibility and hyperfine field were measured in [4]. In accordance with [4], $CuFeSe_2$ is a metal with p type of conductivity and concentration of charge carriers $10^{22}$ $cm^{-3}$ at 300 K. The resistivity temperature dependence has a linear character (Fig. 1). The conductivity of specimens, which was calculated from the data of resistivity changes in the interval ~ 125-71.4 $om^{-1}cm^{-1}$ [4], that is significantly lower then in usual metals.

---

*The conception of magnetopolaron, ferron was introduced by E.L. Nagaev [3]. This conception means that the conductivity electron creates ferromagnetic region around itself by orientation of the spins of neighbor atoms parallel to its spin.

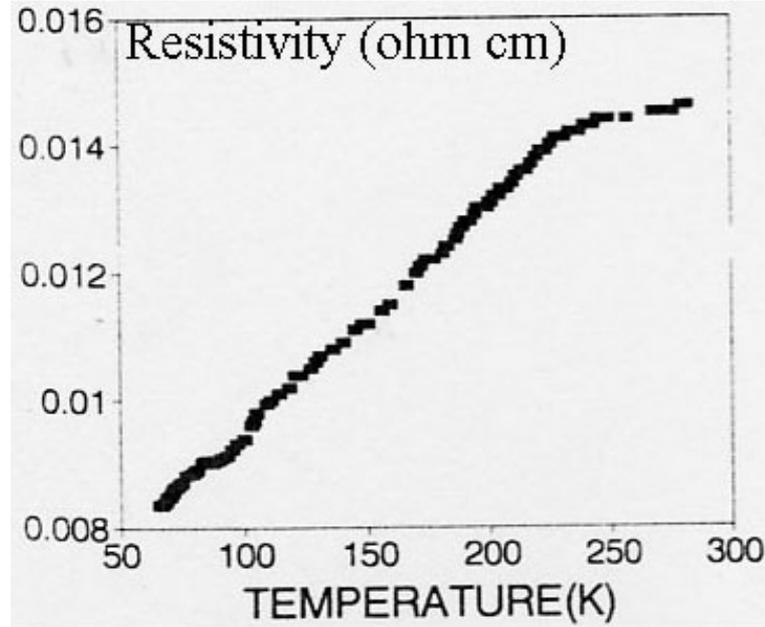

Figure 1. The temperature dependence of resistivity of $CuFeSe_2$

Estimation of the mobility demonstrates, that at 100 K its value is ~0.06 $cm^2V^{-1}s^{-1}$ and is close to those for $CuFeS_2$, which is ~1 $cm^2V^{-1}s^{-1}$ [1], that points out a similar nature of the charge transfer. From the temperature dependence of resistivity $\rho \sim T$ and a constant value of electron density at temperatures higher than $70^0$ K one can deduce, that the conductivity and mobility have the temperature dependence

$$\sigma, \mu \sim 1/T \qquad (1)$$

In accordance with [4] $CuFeSe2$ is paramagnetic in the temperature region $70^0 \leq T \leq 100^0$ K (Fig. 2). Taking into account this fact, one can deduce, that the temperature dependence of mobility (1) indicates the Mott diffusion charge transfer mechanism in this region [6], with ferrons of a small radius and is described by the formula

$$\mu = \frac{ea_0^2}{kT\tau_d\gamma^4} \tag{2}$$

where $a_0$ is the radius of the region occupied by one magnetic atom, $\gamma = R_0/a_0$, $R_0$ is the ferron radius, $\tau_d$ is a spin relaxation constant. All other designations are conventional.

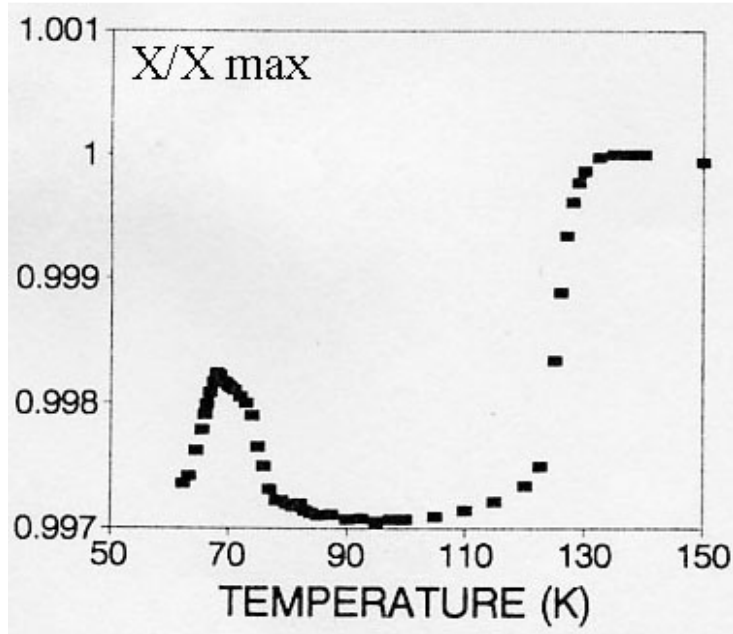

Figure 2. The temperature dependence of magnetic susceptibility $X/X_{max}$ of $CuFeSe_2$

It is shown in [4] that at temperatures below $T \leq 10^0$ K, the resistivity has a constant value, which does not depend on temperature. In spite of $CuFeSe_2$ is a ferrimagnetic, but not a ferromagnetic, one can assume the existence of large radius ferrons in this case. The mobility is constant for the large radius ferrons [6], and is described by the formula

$$\mu(T) \sim \sqrt{\frac{24}{ZS(S+1)}} \frac{ea_0^2}{\hbar} \frac{kT_N}{IS} \tag{3}$$

Here $S$ is the value of the spin of magnetic atom, $Z$ is the number of neighbor magnetic atoms, $a_0$ is a radius of the region, which is occupied by one magnetic atom, $T_N$ is the Neel temperature, $I$ is a value of the exchange integral of the conductivity electron on the magnetic atom. Other designations are conventional.

Since both the mobility and the electron concentration do not depend on temperature, both the specific conductivity and resistivity do not depend on temperature too, that explains their behavior at $T \leq 10$ K. Thus, one can deduce, that the charge transfer in $CuFeSe_2$ in various temperature regions is realized by a small and large radius ferrons that is typical for a large number of magnetic materials.

In conclusion the author wants to thanks N.T. Bagraev for fruitful discussions.

**R E F E R E N C E S:**

1. Kradinova L.V., Polubotko A.M., Popov V.V. *et al.* Novel zero gap compounds, magnetics: $CuFeS_2$ and $CuFeTe_2$ //Semiconductors Science and Technology 1993, V. 8, P. 1616-1619.

2. Vaipolin A.A., Kijaev S.A., Kradinova L.V. *et al.* Investigation of the gapless state in $CuFeTe_2$ // J. Phys: Condensed Matter 1992, V. 4, P. 8035-8039.

3. Nagaev E.L. The ground state and anomalous magnetic moment of conductivity electron in antiferromagnetic semiconductor // JETF Letters 1967, V. 6 P. 484-486. [In Russian].

4. Lamazares J., Gonzalez F., Jaimes E. *et al.* Magnetic, transport, X-ray diffraction and Mossbauer measurements on $CuFeSe_2$ // Journal of Magnetism and Magnetic Materials 1992, V. 104-107, P. 997-998.

5. Woolley J.C., Lamarche A.-M., Lamarche G. *et al.* Low temperature magnetic behaviour of $CuFeSe_2$ from neutron diffraction data // Journal of Magnetism and Magnetic Materials 1996, V. 164, P. 154-162.

6. Kasya T., Janase A., Takeda T. Mobility of a large magnetic polaron// Solid State Communications 1970, V. 8, P. 1551-1553.